\documentclass[letterpaper]{article} % DO NOT CHANGE THIS
\usepackage{aaai24}  % DO NOT CHANGE THIS
\usepackage{times}  % DO NOT CHANGE THIS
\usepackage{helvet}  % DO NOT CHANGE THIS
\usepackage{courier}  % DO NOT CHANGE THIS
\usepackage[hyphens]{url}  % DO NOT CHANGE THIS
\usepackage{graphicx} % DO NOT CHANGE THIS
\usepackage{natbib}  % DO NOT CHANGE THIS AND DO NOT ADD ANY OPTIONS TO IT
\usepackage{caption} % DO NOT CHANGE THIS AND DO NOT ADD ANY OPTIONS TO IT
\usepackage{algorithm}
\usepackage{algorithmic}

\usepackage{newfloat}
\usepackage{listings}
\usepackage{xcolor}
\usepackage{tabularx}

\DeclareCaptionStyle{ruled}{labelfont=normalfont,labelsep=colon,strut=off} % DO NOT CHANGE THIS
\floatstyle{ruled}
\newfloat{listing}{tb}{lst}{}
\floatname{listing}{Listing}
\title{GPT-in-the-Loop: Adaptive Decision-Making for Multiagent Systems}
\author{
    Nathalia Nascimento,
    Paulo Alencar,
    Donald Cowan
}
\affiliations{
    \textsuperscript{}David R. Cheriton School of Computer Science\\
    University of Waterloo (UW)\\
    Waterloo, Canada\\
    \{nmoraesd, palencar, dcowan\} @uwaterloo.ca
}

\begin{document}

%Leveraging Large Language Models in IoT: Pioneering a Self-Adaptive Multiagent Evolution

%to Support Human-Machine Teaming
% LLM-based adaptation 

% \author{
% 	\IEEEauthorblockN{Nathalia Nascimento, Paulo Alencar, Donald Cowan}
% 	\IEEEauthorblockA{\textit{David R. Cheriton School of Computer Science} \\
% 		\textit{University of Waterloo (UW)}\\
% 		Waterloo, Canada \\
% 		\{nmoraesd, palencar, dcowan\} @uwaterloo.ca}
% }

\maketitle

\begin{abstract}
This paper introduces the ``GPT-in-the-loop" approach, a novel method combining the advanced reasoning capabilities of Large Language Models (LLMs) like Generative Pre-trained Transformers (GPT) with multiagent (MAS) systems. Venturing beyond traditional adaptive approaches that generally require long training processes, our framework employs GPT-4 for enhanced problem-solving and explanation skills. Our experimental backdrop is the smart streetlight Internet of Things (IoT) application. Here, agents use sensors, actuators, and neural networks to create an energy-efficient lighting system. By integrating GPT-4, these agents achieve superior decision-making and adaptability without the need for extensive training. We compare this approach with both traditional neuroevolutionary methods and solutions provided by software engineers, underlining the potential of GPT-driven multiagent systems in IoT. Structurally, the paper outlines the incorporation of GPT into the agent-driven Framework for the Internet of Things (FIoT), introduces our proposed GPT-in-the-loop approach, presents comparative results in the IoT context, and concludes with insights and future directions.

% In the realm of multiagent systems, achieving self-adaptation is challenging. This paper introduces a pioneering approach that employs Large Language Models (LLMs), specifically GPT-4, to reshape the evolutionary mechanisms behind self-adaptive agents. Within our framework, each agent, harnesses GPT-4 not only as a decision-making engine but also as a conduit for human-agent interactions, bridging the gap between machine autonomy and human insight. While traditional evolutionary robotics leveraged processes such as neuroevolution for agent optimization, our exploration shifts focus towards the potential of GPT-4 to refine decision-making and human-machine collaboration within dynamic environments. Through our smart street light testbed simulation, agents incorporate the ``GPT-in-the-loop'' model for enhanced reasoning in an Internet of Things (IoT) application scenario. By merging LLMs with IoT, we venture into new realms of agent interaction and problem-solving, aiming to underscore the capabilities LLM-driven multiagent systems bring to the IoT domain.
% ``Adaptive approaches generally require long training processes. " 
%  The continued exploration of these models' reasoning capabilities is indeed intriguing and holds promise for the future of artificial intelligence.
%  These experiments promise to explore novel territories of interaction and problem-solving, thereby pushing the boundaries of what self-adaptive LLM multi-agent systems can achieve.
% %falar do MAPE-K

\end{abstract}


\section{Introduction}
Exploratory investigations are currently underway to harness the reasoning capabilities of Generative Pre-trained Transformers (GPT) for practical applications. Recent studies \cite{richardson2022pushing} \cite{webb2023emergent} \cite{wei2022chain} \cite{huang2023reasoning} indicate that large language models, especially those exceeding 100 billion parameters, are showcasing emergent reasoning abilities. Webb et al. \cite{webb2023emergent} demonstrated that models like GPT-3 might match or even outdo human reasoning in certain tasks—a trajectory GPT-4 seems set to follow. Further supporting this, \cite{wei2022chain} reveals that a ``chain of thought" approach can significantly enhance reasoning in these models, suggesting new methods to utilize their reasoning prowess in real-world scenarios.

Conversely, in the multiagent domain, developing autonomous systems, especially agents that autonomously develop their skills through environment interactions, is an ambitious scientific endeavor \cite{nolfi2022progress}. These agents aim to expand their behavioral repertoire in an open-ended manner. A major thrust is enabling them to employ world models, using common sense knowledge akin to humans, to enhance their performance \cite{nolfi2022progress}. Such knowledge can be gleaned via self-supervised learning, allowing agents to mentally plan and reason. While neuroevolutionary approaches offer potential solutions \cite{almansoori2023evolution,do2017fiot,lan2019evolutionary}, refining neural networks for performance enhancement is time-intensive, costly, and complex, especially in real-time settings with physical agents. In addition to problem-solving skills, the agents should also offer an explanation for their decisions \cite{10.1145/3564240}.

Bridging these two domains, the concept of ``GPT-in-the-loop" emerges as a promising approach. By leveraging the advanced reasoning capabilities of GPT models within the loop of agent decision-making, there's potential to address the challenges in multiagent systems more efficiently. This fusion could harness GPT's inherent adaptability and reasoning prowess, potentially reducing reliance on long training processes that are usual to adaptive approaches \cite{nolfi2022progress}. Inspired by human-in-the-loop approaches \cite{mosqueira2023human}, our proposal defines novel GPT and multiagent system interactions. %We delineate three interaction types: active MAS, where traditional algorithms guide agents and GPT elucidates results, exploring its explainability skills; interactive MAS, fostering tighter collaboration between GPT reasoning and the MAS; and MAS teaching, where GPT guides the MAS adaptation. This work offers examples of each interaction, showcasing one in detail.

Building upon the FIoT framework for adaptable Internet of Things (IoT) applications \cite{do2017fiot}, we incorporate the ``GPT-in-the-loop" methodology. To create self-adaptive IoT agents, FIoT supports the use of different decision-making engines, like neural networks, state machines, and if-else statement; as the use of different adaptative processes, like evolutionary algorithms, backpropagation, and reinforcement learning. FIoT's flexibility in decision-making engines and adaptive processes make it conducive for GPT integration. This flexibility paves the way for GPT to augment reasoning or adaptive functions. For instance, within an interactive MAS setup, GPT can amplify decision-making, aiding agents in outputs and interactions. In MAS teaching, GPT might guide the adaptive process or even dictate the decision-making engine entirely, adjusting agent behaviors based on environmental feedback.

Furthermore, we have applied the GPT-in-the-loop model to smart streetlights, a benchmark IoT application \cite{nascimento2018toward}. In this scenario, agents, equipped with sensors, actuators, and a neural network, evolve to develop a communication system and behavior that optimizes energy while ensuring adequate lighting. As this study \cite{nascimento2018toward} also assessed 14 software engineers' solutions to the same challenge, it allows us to perform a direct comparison between the neuroevolutionary approach, the engineers' solutions, and our GPT-in-the-loop method.

The paper is structured as follows: Section 2 delves into the GPT and FIoT background. Section 3 details our primary contribution, the GPT-in-the-loop approach. Section 4 offers performance results and comparisons within the IoT scenario. We conclude in Section 5.

%In MAS teaching, GPT might either oversee the evolutionary process while agents use basic neural networks, or it could dictate the entire decision-making mechanism, dynamically adjusting to agents' environmental interactions.
%(e.g. creating and adapting the if-else statement to be used by the agents at runtime according to their environment interaction performance). 

 %that we use as our benchmark. 

%and Related Work
\section{Background} \label{sec:background}

\subsection{LLM and GPT}
%LLM- o que ta por tras da estrutura, de forma q permita q ele seja o analyze, plan e o knowledge. 

Large Language Models (LLMs) and Generative Pre-trained Transformers (GPT) are integral parts of AI's Natural Language Processing (NLP) realm. While LLM is a broad category encompassing models that predict word sequences and can be used for various tasks such as text generation and translation, GPT, developed by OpenAI \cite{OpenAI2023}, is a specific LLM type. GPT, renowned for generating text akin to human writing, undergoes extensive pre-training before fine-tuning for specialized tasks. In essence, GPT is a subclass of LLMs, but not all LLMs are GPT models. Other prominent LLM examples include BERT, RoBERTa, and XLNet.

GPT (Generative Pre-trained Transformer) is rooted in the Transformer neural network design \cite{vaswani2017attention,brown2020language}. Representing breakthroughs in natural language processing, GPT, especially in its advanced iterations like GPT-4, utilizes a deep architecture of many layers of these transformers. %This deep configuration equips GPT to interpret and craft human-like text by drawing from its extensive training data. Transformers are pivotal to contemporary strides in NLP due to their prowess in managing sequential datasets, such as written language.
A GPT solution comprises several key components, such as a pre-trained neural network model, a fine-tuning component to improve the model for specific tasks, an inference engine that uses the fine-tuned GPT model to generate responses or predictions (i.e. the inference engine feeds input data into the model and processes the model's output), and data pipeline that handles the flow of data in and out of the model \cite{brown2020language}. 

\subsection{FIoT: Framework for Self-Adaptive IoT Multiagent Systems} \label{sub:FIoT}
The Framework for the Internet of Things (FIoT) \cite{do2017fiot,fiotrepo} is a software framework designed for building control systems for self-operating agents through learning or rule-based methods. Utilizing FIoT results in a Java software element pre-loaded with features for recognizing autonomous entities, assigning control, developing software agents, collecting device data, and ensuring agent-device interactions.

FIoT's features can be customized based on the application's needs. These include: 1) a control unit, ranging from basic if-else conditions to neural networks or preset state machines; 2) a controller adaptation method using techniques like reinforcement learning or genetic algorithms; and 3) a mechanism to evaluate decision-making processes in controlled devices.

There are two primary agents in FIoT: AdaptiveAgent and ObserverAgent.
The former oversees IoT devices and uses the controller for decision-making. Its foundation is the MAPE-K loop \cite{redbooks2004practical}, an esteemed model for enhancing system autonomy. It perceives, acts, and reasons, tailoring outputs based on the chosen decision system. Meanwhile, the ObserverAgent gauges overall agent activity and can refine the control system adopted by IoT agents.

\section{Approach: GPT-in-the-loop} \label{sec:approach}

\begin{figure}[htb!]
	\centering
	\includegraphics[scale=0.45]{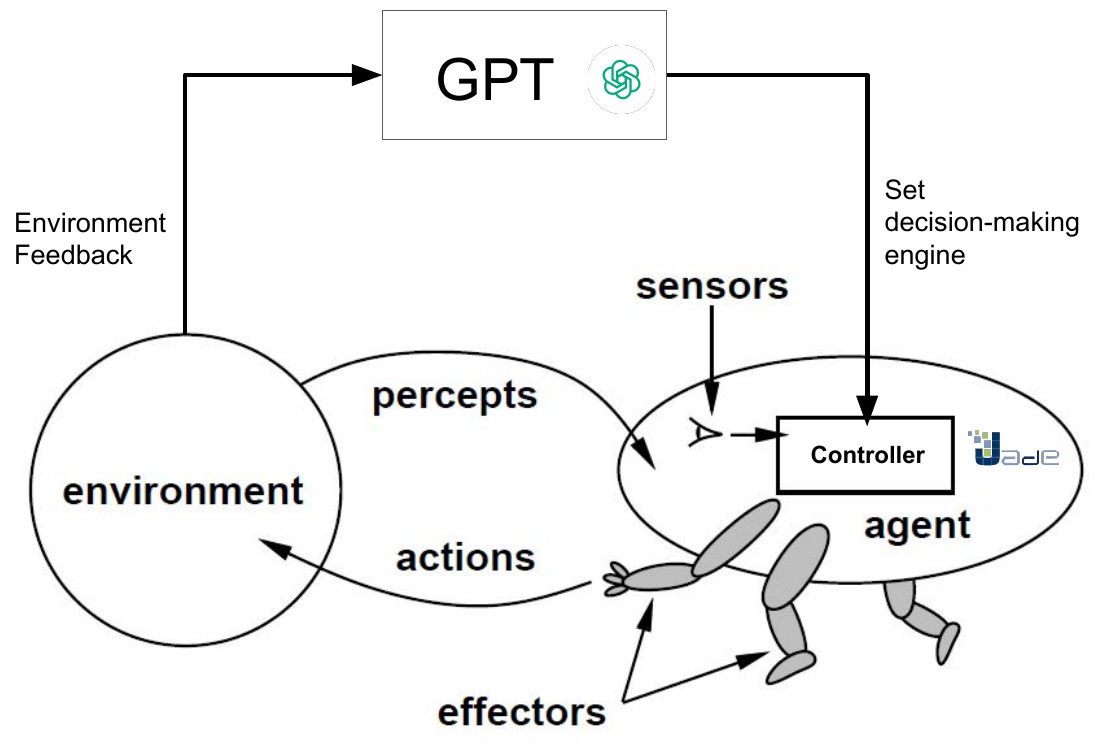} 
	\caption{GPT-in-the-loop: GPT crafts the decision-making engine for the agent, drawing from environmental feedback.}
	\label{fig:approach}
\end{figure}

Drawing inspiration from human-in-the-loop methodologies \cite{mosqueira2023human}, our proposition delineates novel interactions between GPT and multiagent systems. We propose three main interaction modes: 

\begin{itemize}
    \item \textbf{Active MAS:} Traditional algorithms drive agents while GPT clarifies outcomes.
    \item \textbf{Interactive MAS:} This encourages a more integrated collaboration between the GPT's reasoning and the MAS, which has been our primary focus in this work as depicted in Figure \ref{fig:approach}.
    \item \textbf{MAS Teaching:} Here, the GPT directs the MAS adaptation.
\end{itemize}

In the interactive MAS model, GPT shapes the decision-making engine of the agent. This engine processes inputs, generates outcomes, and influences the manner in which the agent engages with its environment, which in turn impacts application performance. Feedback from these engagements can re-engage the GPT, leading to refinements in agent behaviors. 

While the interactive MAS mode stands at the heart of our research, we chose to integrate GPT with FIoT. This framework paves the way for probing diverse interaction forms. It permits a complete overhaul of the IoT agents' decision-making engine or, alternatively, steers the evolution/training process orchestrated by the ObserverAgent.

\begin{figure}[ht!]
	\centering
	\includegraphics[scale=0.47]{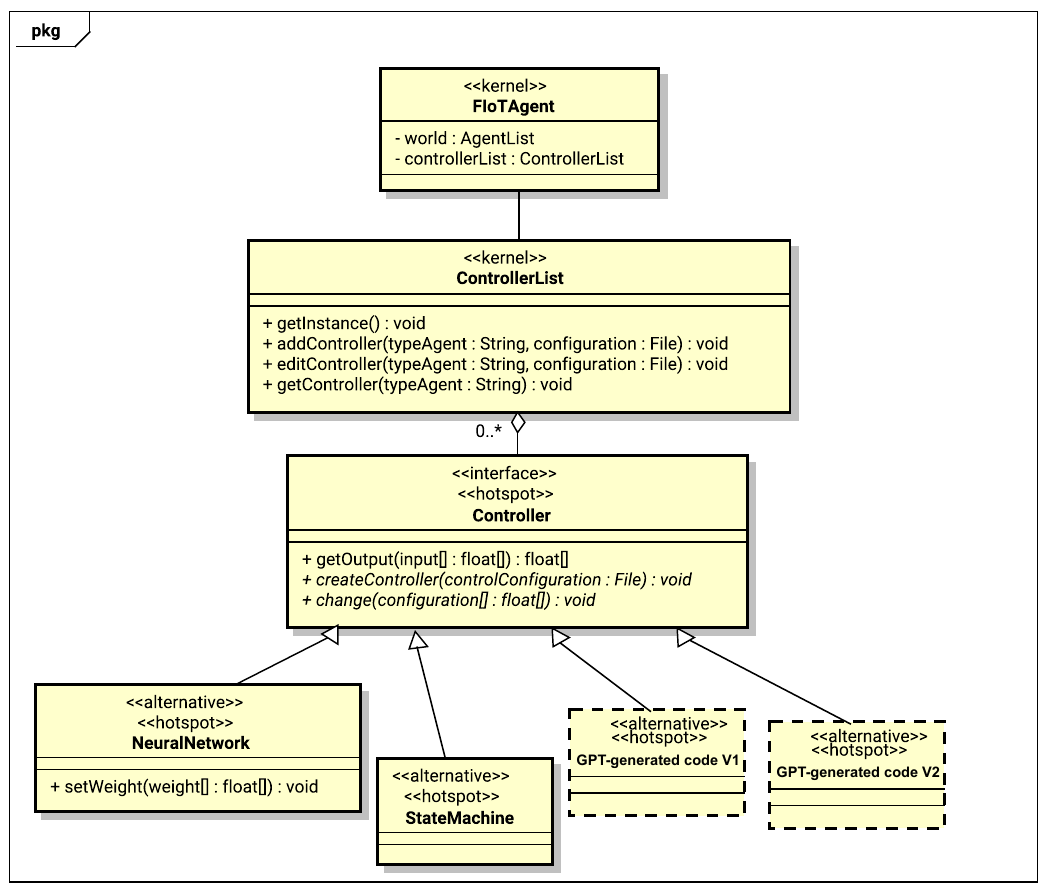} 
	\caption{Augmenting FIoT to empower agents with decision-making abilities using GPT-crafted code.}
	\label{fig:behavioruml1}
\end{figure}

\begin{figure*}[ht!]
	\centering
	\includegraphics[scale=0.85]{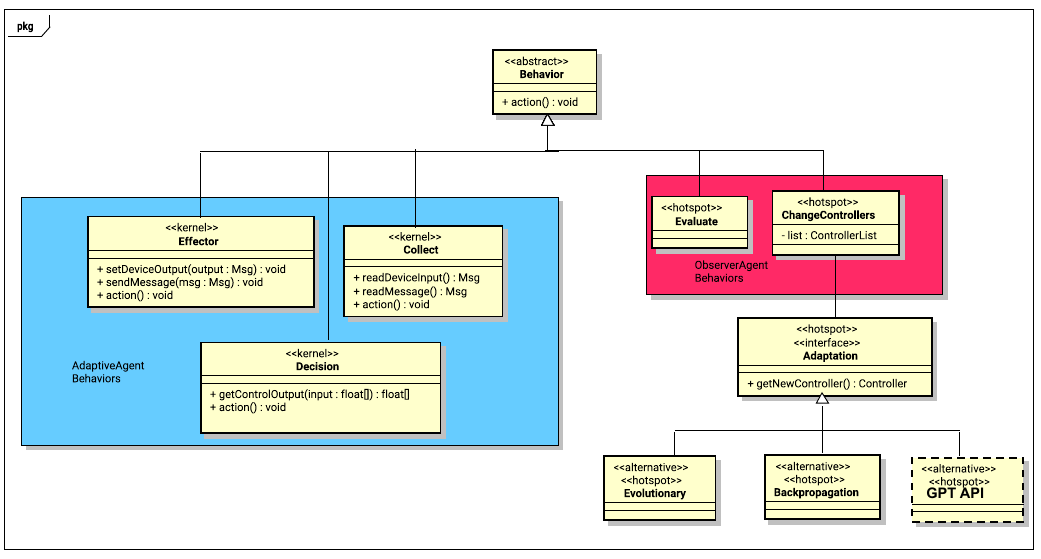} 
	\caption{Elevating FIoT to incorporate GPT as a potential adaptive strategy for the Observer Agent.}
	\label{fig:behavioruml2}
\end{figure*}

Figures \ref{fig:behavioruml1} and \ref{fig:behavioruml2} illustrate the seamless extension of FIoT to accommodate the GPT-in-the-loop model, tapping into both the AdaptiveAgent's controller and the ObserverAgent's adaptation process. Notably, both the decision-making controller and the adaptive procedure are flexible points at the framework. This allows for varied runtime implementations, so long as class signatures (parameters, inputs, and outputs) remain consistent. For instance, environmental feedback can prompt GPT to craft a new controller for agents.

\section{Application Scenario: Smart Streetlights}
\label{section:app}
In our experiment, we replicated the streetlight scenario from \cite{nascimento2018toward} using the FIoT framework. The goal was to create autonomous streetlights balancing energy conservation with effective illumination, ensuring individuals could navigate their paths seamlessly. These streetlights, equipped with sensors and communicative tools, had three core functions: data collection, decision-making, and action execution. The focus of this experiment was on the decision-making aspect. 

The original study utilized a three-layer neural network, evolved through a genetic algorithm, to automate the streetlights' decision rules. Software engineers also tackled the challenge, developing decision-making solutions. They were presented with the same simulated scenario, facilitating a comparison of human-devised solutions with the automated neural network method. Subsequently, these solutions were tested in an expanded environment. This second phase aimed to assess whether the decision-making module, originally designed for the first scenario, could be effectively reused in a different environment.

Incorporating the GPT-in-the-loop methodology, and paralleling the strategy in \cite{nascimento2018toward}, GPT engaged with the primary scenario until it derived a solution surpassing a fitness score of 62 (we set it based on the best fitness value presented in \cite{nascimento2018toward}). This derived decision mechanism was then trialed in the expansive environment. Conclusively, we set the GPT-in-the-loop results as a benchmark, juxtaposing them against the top solutions from the neuroevolutionary algorithm, the best software engineer participant, and GPT's own solution.

To facilitate a clear comparison between the two methods, Table~\ref{table:case2} showcases the application of the Streetlight Control case study using a neuroevolutionary approach, highlighting the flexible points of the FIoT framework. Conversely, Table~\ref{table:case3} delineates the implementation of the Streetlight Control application through the GPT-in-the-loop-based approach, capitalizing on the adaptability of the FIoT framework. Both tables aim to provide a foundation for evaluating the efficacy of each solution within the same application context.

\begin{table}[htb!]
	\centering
	\caption{Implementing FIoT flexible points to synthesize streetlight controllers using an ML-based approach \cite{nascimento2018toward}.}
	\begin{tabular}{|l|l|}
		\hline
		\textbf{FIoT Framework}    & \textbf{Light Control Application}                                                                                                                                                                                                                                  \\ \hline
		Controller             & Three Layer Neural Network                                                                                                                                                                                                                           \\ \hline
		Making Evaluation     & \begin{tabular}[c]{@{}l@{}}Collective Fitness Evaluation: \\the solution is evaluated \\based on the energy \\consumption, the number of \\people that finished their \\routes after the \\simulation ends, and the \\total time spent by people \\to move during their trip\end{tabular}  \\ \hline
		\begin{tabular}[c]{@{}l@{}}
			Controller Adaptation 
		\end{tabular} 
		& \begin{tabular}[c]{@{}l@{}}Evolutionary Algorithm: \\Generate a pool of \\candidates  to represent the\\neural network parameters\end{tabular}                                                                                                                  \\ \hline
	\end{tabular}
	\label{table:case2}
\end{table}

\begin{table}[htb!]
	\centering
	\caption{Detailed implementation of FIoT flexible points to synthesize streetlight controllers using the GPT-in-the-loop approach.}
	\begin{tabular}{|l|l|}
		\hline
		\textbf{FIoT Framework}    & \textbf{Light Control Application}                                                                                                                                                                                                                                       \\ \hline
		Controller                 & \begin{tabular}[c]{@{}l@{}}GPT-based Decision Engine: \\Use if-else statement controllers \\optimized for the given \\scenario's constraints and goals\end{tabular}                                                                                       \\ \hline
		Making Evaluation     & \begin{tabular}[c]{@{}l@{}}Iterative Fitness Evaluation: \\the solution iterates until it\\exceeds a fitness score of 62, \\evaluating based on energy \\consumption, the number of \\people that complete routes, \\and the cumulative time of \\people's journeys\end{tabular}         \\ \hline
		\begin{tabular}[c]{@{}l@{}}
			Controller Adaptation 
		\end{tabular} 
		& \begin{tabular}[c]{@{}l@{}}GPT-in-the-loop: \\GPT engages in \\interactive loops, refining \\its if-else controllers based on \\environment feedback until\\ the desired fitness \\level is reached\end{tabular} \\ \hline
  % \begin{tabular}[c]{@{}l@{}}GPT-in-the-loop: \\GPT interacts with the first \\scenario and refines its solutions \\based on feedback loops until \\achieving desired fitness\end{tabular}                                                                                                                   \\ \hline
	\end{tabular}
	\label{table:case3}
\end{table}

\subsection{Results and Discussion}

The GPT-in-the-loop approach required three iterations to reach a fitness score of 62 in the first scenario. Comparatively, the original evolutionary approach underwent 200 generations, with each generation undergoing 50 interactions with the environment. This section details the solutions GPT proposed for the streetlight controllers in each iteration, elucidating the adaptive changes influenced by GPT's explanations. Ultimately, we compared the outcomes of this innovative method against those of neuroevolution and human-generated solutions.

In the initial iteration, the problem description presented to the software engineers in \cite{nascimento2018toward} served as the prompt. For every iteration, GPT generated an `if-else statement' controller accompanied by a rationale for the decision.

The code, as shown in Figure \ref{table:codev1}, emerged from the first iteration. The foundational idea behind this solution was: 1) Ambient Light Detection: It's the sole strategy highlighting that poles will perpetually be in ``listening" mode, regardless of external lighting conditions; 2) Wireless Communication: Stresses the significance of interaction with the preceding pole and adjusting illumination based on that communication; 3) Energy Conservation: Sets a distinct condition whereby, in the absence of motion detection for a predetermined cycle count, the light's intensity diminishes.

\begin{figure}[ht!]
\centering
\caption{GPT's decision-making engine from the first iteration.}
\label{table:codev1}
%\begin{tabular}{|c|}
%\hline
\begin{minipage}{\linewidth}
\begin{lstlisting}[basicstyle=\tiny\ttfamily,numbers=none]
@Override
public double[] getOutput(double[] input) {
    // Input variables
    double lightingSensor = this.getLightingSensor();
    double motionSensor = this.getMotionSensor();
    double wirelessReceiver = (this.getPreviousListeningDecision() == 1.0) ? this.getWirelessReceiver() : 0.0;  // If not listening, ignore the value.
    // Decision variables
    double lightDecision = 0.0;  // By default, we keep it off
    double wirelessTransmit = 0.0;  // By default, we don't transmit anything
    double listeningDecision = 1.0;  // By default, we always listen to the surrounding posts
    // If it's day, we keep the light off
    if (lightingSensor == 1.0) {
        lightDecision = 0.0;
    } else {
        // If we detect movement
        if (motionSensor == 1.0) {
            lightDecision = 1.0;
            wirelessTransmit = 1.0;  // We alert the surrounding posts about the detected movement
        }
        // If the neighboring post detected movement and we are listening
        else if (wirelessReceiver > 0.0) {
            lightDecision = 0.5;  // We partially light up to prepare the path
        }
    }
    // Action decisions
    this.setLightDecision(lightDecision);
    this.setWirelessTransmitter(wirelessTransmit);
    this.setListeningDecision(listeningDecision);
    return this.actuador;
}
\end{lstlisting}
\end{minipage} \\
%\hline
%\end{tabular}
\end{figure}

The first iteration resulted in an average fitness of 29.49. Given its inadequacy, GPT was re-engaged to produce an enhanced solution. For the second iteration, the environment's outputs, such as energy consumption and trip time, were incorporated into the GPT prompt. The code from the second iteration is showcased in Figure \ref{table:codev2}. Contrary to the first solution, which emphasized persistent ambient light listening, previous pole communication, and energy preservation, the second solution accentuates anticipatory lighting influenced by adjacent movements and subdued illumination during extreme darkness.

\begin{figure}[ht!]
\centering
\caption{GPT's decision-making engine from the second iteration (input and output settings excluded).}
\label{table:codev2}
%\begin{tabular}{|c|}
%\hline
\begin{minipage}{\linewidth}
\begin{lstlisting}[basicstyle=\scriptsize\ttfamily,numbers=none]
@Override
public double[] getOutput(double[] input) {
   /*...*/
    double lightDecision = 0.0;
    double wirelessTransmit = 0.0;
    double listeningDecision = 1.0;  // Always listening to optimize the person's path
    // If motion is detected
    if (motionSensor == 1.0) {
        lightDecision = 1.0;  // Turn on fully for the passerby
        wirelessTransmit = 1.0;  // Notify the nearby posts to prepare the path
    // If a neighboring post reports movement
    } else if (wirelessReceiver == 1.0) {
        lightDecision = 0.5;  // Partially light up to anticipate the passerby's arrival
    // Energy saving without compromising safety
    } else if (lightingSensor == 0.0) {
        lightDecision = 0.5;  // Keep partially lit if it's too dark, to ensure some level of safety
    }
    /*...*/
    return this.actuador;
}
\end{lstlisting}
\end{minipage} \\
%\hline
%\end{tabular}
\end{figure}

The second iteration notably improved the fitness value, achieving 61.2. However, since this score remained below the target, GPT produced another controller, depicted in Figure \ref{table:codev3}. Unlike the solution from the second iteration, this new approach concentrates on curtailing data transmissions by sending signals exclusively upon initial motion detection, determining listening necessity based on present light status, and integrating adaptive dimming.

\begin{figure}[!ht]
\centering
\caption{GPT's decision-making engine from the third iteration (input and output settings excluded).}
\label{table:codev3}
\begin{minipage}{\linewidth}
\begin{lstlisting}[basicstyle=\scriptsize\ttfamily,numbers=none]
@Override
    public double[] getOutput(double[] input) {
        /*...*/
        if (motionSensor == 1.0) {
            lightDecision = 1.0;
            if (this.getPreviousListeningDecision() != 1.0) {
                wirelessTransmit = 1.0; // Transmit motion detection once
                listeningDecision = 0.0; // No need to listen when we already detect motion
            }
        } else if (wirelessReceiver == 1.0) {
            lightDecision = 0.5;
            if (this.getPreviousListeningDecision() != 1.0) {
                wirelessTransmit = 0.5; // Inform neighbors about anticipatory lighting
            }
        } else if (lightingSensor == 0.0) {
            lightDecision = 0.5;
        }
        /*...*/
        return this.actuador;
    }
\end{lstlisting}
\end{minipage} \\
\end{figure}

The integration of environmental feedback into successive GPT iterations consistently led to performance improvements. Table \ref{table:result1} displays the fitness outcomes across three iterations for scenario 1, whereas Table \ref{table:result2} presents the outcomes when the decision algorithms were implemented in a more intricate environment for scenario 2. In both scenarios, the third solution proposed by GPT outperformed its predecessors. When juxtaposed with the optimal outcome derived from the neuroevolution method, the solution offered by GPT achieved a superior fitness score in both scenarios. In scenario 1, one participant managed to devise a solution with a slightly better fitness score than that of GPT's. However, this solution faltered in the second scenario. Evaluating GPT's performance against the most successful participant-driven solutions in the second scenario, GPT's solution was unrivaled.

\begin{table*}[!ht]
\centering
\caption{Performance comparison of GPT iterations, best neuroevolution solution, and best participant's solution in the first scenario.}
     \label{table:result1}
\begin{tabular}{|c|c|c|c|c|}
\hline
\textbf{Solution}              & \textbf{Energy} & \textbf{People} & \textbf{TotalFTrip} & \textbf{Fitness} \\ \hline
GPT (iteration 1)              & 4.03            & 66.66           & 59.25               & 29.49            \\ \hline
GPT (iteration 2)              & 15.02           & 100             & 54.62               & 61.2             \\ \hline
GPT (iteration 3)              & 11.92           & 100             & 54.62               & 62.44            \\ \hline
Best neuroevolution's solution & 8.1             & 100             & 62.03               & 59.53            \\ \hline
Best participant's solution    & 9.46            & 100             & 55.55               & 62.88            \\ \hline
\end{tabular}
\end{table*}

\begin{table*}[!htb]
\centering
\caption{Performance comparison of GPT iterations, best neuroevolution solution, and best participant's solution in the second scenario.}
     \label{table:result2}
\begin{tabular}{|c|c|c|c|c|}
\hline
\textbf{Solution}              & \textbf{Energy} & \textbf{People} & \textbf{TotalFTrip} & \textbf{Fitness} \\ \hline
GPT (iteration 1)              & 2.08            & 66.66    & 48.51      & 36.72            \\ \hline
GPT (iteration 2)              & 11.29          & 100             & 41.10       & 70.81     \\ \hline
GPT (iteration 3)              & 9.76    & 100             & 41.10              & 71.42           \\ \hline
Best neuroevolution's solution & 8.46            & 100             & 46.29               & 68.83            \\ \hline
Best participant's solution    & 50.52           & 100             & 38.14               & 56.9             \\ \hline
\end{tabular}
\end{table*}

In the quest for optimized streetlight controllers, GPT's iterative approach showcased notable adaptability and improvement. The model's ability to integrate environment feedback between iterations culminated in solutions competitive with human and neuroevolution strategies. Moreover, GPT's intrinsic explainability, as evident in its generated `if-else statements' and accompanying rationale, offers valuable insights for users, bridging the gap between automation and human understanding.

\section{Conclusion and Future Horizons} \label{sec:results}

The synergy between Large Language Models (LLMs) like GPT-4 and multiagent systems promises to redefine the boundaries of autonomous interactions and adaptability.  Our research underscores the compelling advantages of this integration. The GPT-in-the-loop methodology exemplifies how problem-solving abilities can be significantly enhanced in a dynamic setting. When LLMs are incorporated into agents, it catalyzes a two-fold benefit: a supercharged reasoning mechanism for each agent and a more efficient communication process across a diverse multi-agent landscape.

Furthermore, GPT's unique ability to elucidate its decision-making process brings a new dimension of transparency. This clarity not only strengthens confidence in the system's actions but also paves the way for a deeper understanding of intricate decisions.

Nonetheless, this promising integration is met with inherent challenges. From the substantial computational needs of LLMs to the subtleties surrounding their decisions and looming ethical considerations, there's a clear call for meticulous evaluation. The forward-looking vision of agents dynamically leveraging a cloud-hosted GPT to optimize their actions in real-time is undeniably ambitious. To fully materialize this vision, further research and exploration are essential, especially in leveraging GPT-in-the-loop to enhance diverse GPT-Multiagent interactions.

\subsection{Exploring Further Configurations for the GPT-in-the-loop}\label{sub:exploring}

The premise of ``GPT-in-the-loop" holds tremendous potential in the realm of multiagent systems, leveraging the sophisticated reasoning capabilities of GPT models directly into agent decision-making processes. Given the inspiration drawn from the human-in-the-loop approaches \cite{mosqueira2023human}, our roadmap defines diverse GPT and multiagent system interactions, which can be expanded in several directions:

\begin{enumerate}
    \item Active MAS Involvement: A scenario wherein traditional algorithms guide the multiagent systems, and GPT steps in to provide clarity and interpretation of results. This interaction mode mainly draws on GPT's unparalleled explainability prowess, making complex decisions more transparent and comprehensible.
    \begin{enumerate}
        \item  GPT as a Decentralized Decision Engine: A promising direction is to use GPT as the primary decision-maker for each agent. Instead of one general reasoning mechanism for all agents, envision each agent having its personalized GPT. This approach allows agents to make context-specific decisions in real-time, drawing from GPT's vast knowledge to address their unique situations.
    \end{enumerate}
    \item Interactive MAS Integration: This model envisions a more intimate alliance between GPT's reasoning faculties and the multiagent system. Here, there's a bidirectional flow of information and decisions, ensuring that both GPT and MAS evolve and adapt symbiotically.
    \item MAS Teaching: GPT's role as a tutor or mentor to multiagent systems. GPT could oversee, instruct, and guide the adaptation process of MAS. 
\end{enumerate}

\subsection{Enhancing Human Engagement in the Loop}

While the human element remains foundational, especially in shaping the initial system prompt or documentation, the potential for a more intertwined human-machine partnership exists.

\begin{enumerate}
\item Direct Influence: Encouraging humans to directly shape agent behaviors is key. An intuitive interface could enable users to propose behaviors, pinpoint overarching goals, or lay out specific parameters. This merges human intuition with technological prowess, targeting the best results for agents.
\item Feedback Mechanism: It's beneficial when agents offer clear summaries of their decisions, from data analysis to behavioral tweaks. Such transparency strengthens trust, offers clarity, and provides avenues for system enhancements based on human feedback.
\item Making Sense of Complexity: Even though adaptive systems are complex by design, demystifying their workings is essential. Translating intricate operations into comprehensible language paves the way for enhanced human-machine interactions.
\end{enumerate}

\subsection{Diversifying Application Scenarios:}
Venturing beyond our preliminary framework, our ambition is to validate the GPT-in-the-loop approach in a spectrum of applications, especially when integrated with realistic robotics frameworks like Evorobot \cite{evorobotpy2} and Webots \cite{michel2004cyberbotics}. Such platforms enable the deployment of neural networks sculpted by evolutionary techniques.

The domain of evolutionary robotics unravels complex challenges, a notable one being the food foraging task \cite{pontes2022towards}. Here, agents are tasked with distinguishing nourishing food sources from harmful ones, adeptly navigating environmental intricacies for optimal survival. In this setup, agents traverse a dynamic landscape, reliant on a singular light sensor, to ascertain the edibility of proximate food. Represented in alternating colors of black and white, the safety of the food keeps shifting, mandating constant adaptability. Agents face a binary choice: to consume or avoid the food, within a given time frame.

\begin{figure}[htb!]
	\centering
	\includegraphics[scale=0.42]{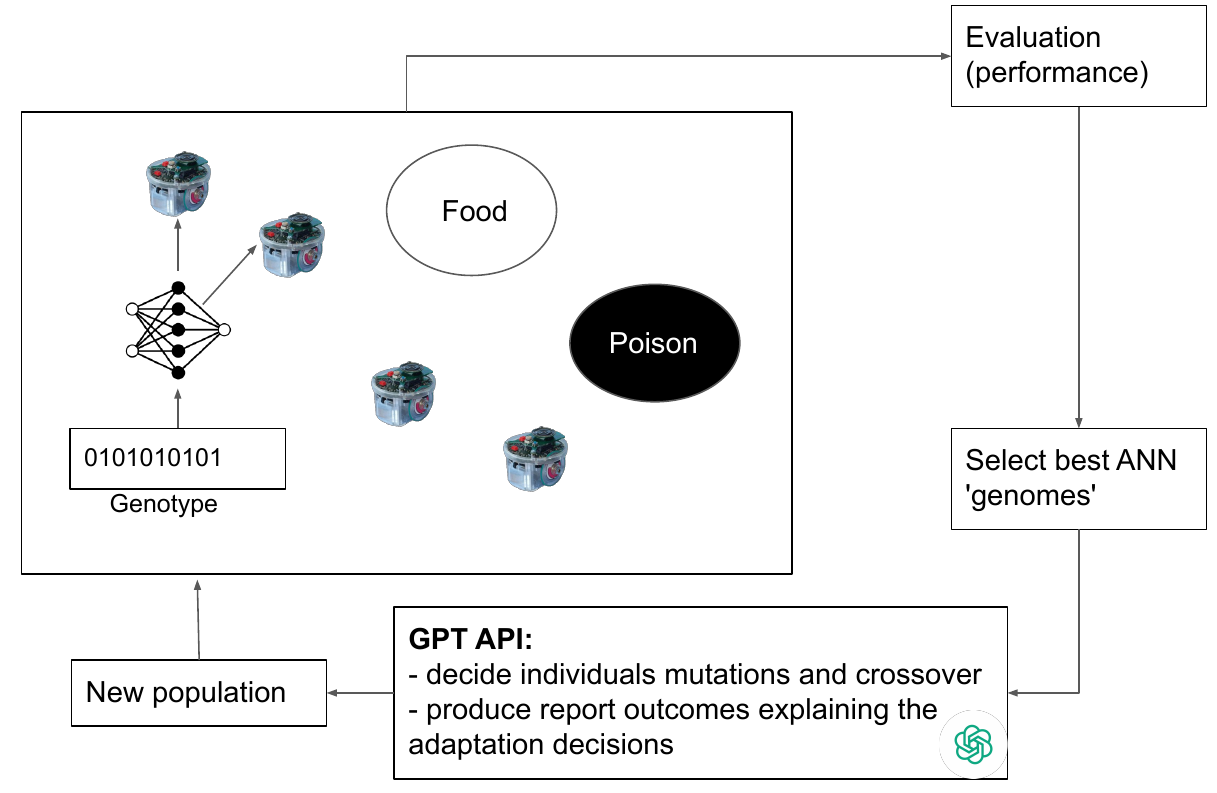} 
	\caption{GPT-in-the-loop: GPT supporting the evolutionary process.}
	\label{fig:future}
\end{figure}

Figure \ref{fig:future} depicts our conceptualization of GPT-in-the-loop within a distinct application setting, accounting for an alternative MAS interaction paradigm. Here, the graphic portrays a MAS teaching interaction: while agents predominantly adhere to a conventional evolutionary path, GPT plays a supportive role in their evolution.

\subsubsection{Evolutionary GPT Engagement:} Embedding GPT within the evolutionary paradigm offers captivating prospects. GPT, transcending its observational role, can proactively shape the evolutionary trajectory. This encompasses guiding individual selection, fine-tuning genetic algorithms, and pinpointing ideal neural network configurations. Incorporating GPT's analytical prowess with evolutionary strategies could potentially evolve solutions that are not only optimal but also explainable.

Integrating LLMs into such narratives exhibits significant potential. With the GPT-in-the-loop approach, we're amplifying agent adaptability and delving deep into the multifaceted GPT-MAS interactions delineated in subsection \ref{sub:exploring}. This synergy might herald a transformative shift in the adaptability and prowess of future robotic agents.

\subsection{Diversifying LLM Choices:} While we centered on GPT-4, many other LLMs exist with unique capabilities. Exploring these options and creating clear evaluation standards might lead to even more effective multiagent strategies.

\subsection{Addressing the Black-Box Concern:} GPT-4 remains proprietary and opaque despite its explanatory capabilities. To ensure trust and safety, there's imperative to decode its operational logic, facilitating rigorous testing and risk-mitigation strategies.

\section*{Acknowledgment}
This work was supported by the Natural Sciences and Engineering Research Council of Canada (NSERC), and the Centre for Community Mapping (COMAP).

\bibliography{references,references-aai}
\end{document}